\documentclass[aps,prb,twocolumn,floatfix]{revtex4}

\usepackage{graphicx}
\usepackage{dcolumn}
\usepackage{bm}
\usepackage{times}
\usepackage{color}

\begin{document}
\bibliographystyle{apsrev}
\title{Large linear magnetoresistance in a new Dirac material BaMnBi$_2$}
\author{Yi-Yan Wang}\thanks{These authors contribute equally to this work}
\author{Qiao-He Yu}\thanks{These authors contribute equally to this work}
\author{Tian-Long Xia}\email{tlxia@ruc.edu.cn}
\affiliation{Department of Physics, Beijing Key Laboratory of Opto-electronic Functional Materials $\&$ Micro-nano Devices, Renmin University of China, Beijing 100872, P. R. China}
\date{\today}
\begin{abstract}
We report the synthesis of high quality single crystals of BaMnBi$_2$ and investigate the transport properties of the samples. The Hall data reveals electron-type carriers and a mobility $\mu(5K)=1500cm^2/Vs$. The temperature dependence of magnetization displays behavior that is different from CaMnBi$_2$ or SrMnBi$_2$, which suggests the possible different magnetic structure of BaMnBi$_2$. Angle-dependent magnetoresistance reveals the quasi-two-dimensional Fermi surface. A crossover from semiclassical MR$\sim$H$^2$ dependence in low field to MR$\sim$H dependence in high field is observed in transverse magnetoresistance. Our results indicate the anisotropic Dirac fermion states in BaMnBi$_2$.
\end{abstract}
\maketitle
\setlength{\parindent}{1em}
\section{Introduction}
\indent Dirac materials is a group of compounds whose low-energy excitations behave as massless Dirac particles\cite{wehling2014dirac}. Recent years, a variety of Dirac materials have been discovered, such as graphene\cite{RevModPhys.81.109}, 3D topological insulators Bi$_{1-x}$Sb$_x$, Bi$_2$Se$_3$ and Bi$_2$Te$_3$\cite{RevModPhys.82.3045,RevModPhys.83.1057}, d-wave cuprate superconductors\cite{RevModPhys.72.969}, Na$_3$Bi\cite{liu2014discovery} and Cd$_3$As$_2$\cite{PhysRevLett.113.027603,neupane2014observation,liang2015ultrahigh,PhysRevLett.113.246402} \emph{et al}. The energy spectrum of Dirac materials exhibits linear behavior and can be described by relativistic Dirac equation. More interestingly, when the time reversal symmetry or space inversion symmetry is broken, Dirac semimetals (DSM) develop into Weyl semimetals (WSM), whose extraordinary properties such as Fermi arc and chiral anomaly have attracted great attention\cite{PhysRevB.83.205101,PhysRevX.5.031013,xu2015discovery,lv2015observation,PhysRevX.5.031023,yang2015weyl,huang2015weyl,xu2015experimental,PhysRevX.5.011029,liu2016evolution,shekhar2015extremely,shekhar2015large,yang2015chiral,PhysRevB.92.115428}
. One interesting consequence of the linear energy dispersion is quantum transport phenomena. For Dirac materials, a moderate magnetic field can compel all carriers occupy the lowest Landau level and lead to the quantum linear magnetoresistance (MR)\cite{zhang2005experimental,miller2009observing}.

\indent In Dirac materials, there is a class of ternary 112 type compounds such as CaMnBi$_2$\cite{PhysRevB.85.041101,he2012giant,feng2014strong,PhysRevB.87.245104}, SrMnBi$_2$\cite{PhysRevLett.107.126402,PhysRevB.84.220401,PhysRevB.84.064428,PhysRevB.90.075120}, EuMnBi$_2$\cite{masuda2016quantum,PhysRevB.90.075109} and LaAgBi$_2$\cite{PhysRevB.87.235101}, which have been researched deeply. More recently, time-reversal symmetry breaking Weyl state has been discovered in YbMnBi$_2$\cite{borisenko2015time}, and Sr$_{1-y}$Mn$_{1-z}$Sb$_2$\cite{liu2015discovery} is also identified as a promising candidate of WSM. Among these materials, the Bi/Sb square net is a common feature and has been considered as the platform that hosts Dirac/Weyl fermions. Therefore, it is of considerable interest to explore new materials which have similar structure and further study their physics properties.

\indent In this work, we have successfully synthesized the single crystals of BaMnBi$_2$ and investigated the transport properties of BaMnBi$_2$. Hall resistivity shows  carriers in BaMnBi$_2$ is electron type. Magnetic property measurement indicates the magnetic structure of BaMnBi$_2$ may be slightly different from that of SrMnBi$_2$ or CaMnBi$_2$. Angle-dependent MR implies the anisotropic quasi-two-dimensional Fermi surface in BaMnBi$_2$. The in-plane MR displays a crossover from semiclassical quadratic field dependence to linear field dependence with the increase of magnetic field. The linear MR indicates the possible existence of Dirac fermions in this material. It is worthy to do ARPES experiments to check whether BaMnBi$_2$ is a Weyl semimetal.

\section{experimental and crystal structure}
\indent Single crystals of BaMnBi$_2$ were grown from Bi flux. The mixtures of Ba, Mn and Bi were placed in a crucible and sealed in a quartz tube with a ratio of Ba:Mn:Bi=1:1:6. The quartz tube was heated to 1180$^0$C in 60 h, held there for 30 h, and cooled to 370$^0$C at a rate of 3$^0$C/h, and then the excess Bi-flux was removed by decanting. Elemental analysis was performed using energy dispersive X-ray spectroscopy (EDS, Oxford X-Max 50). The determined atomic proportion was consistent with the composition of BaMnBi$_2$ within instrumental error. Single crystal X-ray diffraction (XRD) pattern was collected from a Bruker D8 Advance X-ray diffractometer using Cu K$_{\alpha}$ radiation. Resistivity measurements were performed on a Quantum Design physical property measurement system (QD PPMS-14T) and the magnetic properties were measured with vibrating sample magnetometer (VSM) option.

\indent BaMnBi$_2$ is isostructural with SrMnBi$_2$. As shown in Fig.1(a), the crystal structure of BaMnBi$_2$ is comprised of alternating MnBi and BaBi layers\cite{cordier1977darstellung}. In the MnBi layer, each Mn atom is surrounded by four Bi atoms, which form the MnBi$_4$ tetrahedra. In the BaBi layer, Bi atoms are separated and form a square net (highlighted by red atoms). Fig.1(c) shows the X-ray diffraction pattern of a BaMnBi$_2$ single crystal on the (00l) plane. The inset is an optical image of a representative crystal.

\begin{figure}[htbp]
\centering
\includegraphics[width=0.48\textwidth]{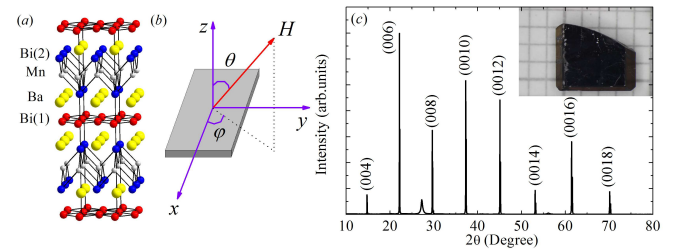}
\caption{(Color online) (a) The crystal structure of BaMnBi$_2$. (b) The definition of polar angle $\theta$ and azimuthal angle $\varphi$. (c)Single crystal X-ray diffraction pattern of a BaMnBi$_2$ crystal, showing only the (00l) reflections. The inset is an image of a typical single crystal with a scale of 4 mm.}
\end{figure}

\begin{figure}[htbp]
\centering
\includegraphics[width=0.48\textwidth]{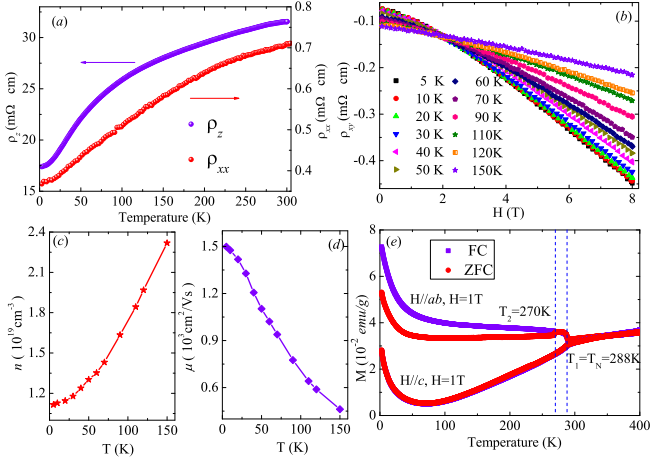}
\caption{(Color online) (a) Temperature dependence of in-plane resistivity $\rho_{xx}(T)$ and out-plane resistivity $\rho_z(T)$. (b) Hall resistivity $\rho_{xy}$ vs magnetic field H with the temperature T from 5K to 150K. (c)-(d) Temperature dependence of carrier concentration and mobility, respectively. (e) Magnetization versus temperature with the magnetic field H=1T applied parallel to ab plane (up) and c axis (down) under ZFC (red) and FC (violet) conditions. The dash lines denote two anomalous temperatures.}
\end{figure}

\section{Results and discussions}
\indent Both the in-plane resistivity $\rho_{xx}(T)$ and out-plane resistivity $\rho_z(T)$ exhibit a simple metallic behavior as shown in Fig.2(a). The out-plane resistivity is nearly 50 times of in-plane resistivity in the range 2.5-300K. Such a significant anisotropy suggests quasi-2D electronic band structure in BaMnBi$_2$. Fig.2(b) plots the magnetic field dependence of Hall resistivity $\rho_{xy}(H)$ measured at various temperatures. The negative slope of $\rho_{xy}$ suggests that the dominant charge carriers in BaMnBi$_2$ are electrons. Single band model has been employed to analyze the Hall effect data. The carrier concentration and carrier mobility are shown in Fig.2(c) and (d) respectively. The carrier concentration is given by $n$=$1/eR_H$, and the Hall coefficient $R_H$ is obtained by $R_H$=$\rho_{xy}(8T)/H$. At 5K, the carrier concentration is $1.1\times10^{19}cm^{-3}$. We calculate the carrier mobility by $\mu$=$1/en\rho_{xx}(0 T)$. With temperature reduced, the mobility becomes larger and reaches 1500$cm^2/Vs$ at 5K.

\indent Fig.2(e) presents the temperature dependence of magnetization of BaMnBi$_2$ measured under zero field cooling (ZFC) and field cooling (FC) conditions with an applied field H=1T parallel to ab plane and c axis respectively. Similar to CaMnBi$_2$ and SrMnBi$_2$\cite{PhysRevB.90.075120,PhysRevB.84.064428}, there are two anomalous temperatures in BaMnBi$_2$. T$_1$=T$_N$=288K is the temperature of antiferromagnetic transition, below T$_1$ the magnetization exhibits strong anisotropy, the linear temperature dependence of magnetization above T$_1$ indicates strong antiferromagnetic correlations\cite{zhang2009universal}. There is a ZFC-FC splitting for magnetic field parallel to ab plane when $T<T_2$=270K. This is quite different from CaMnBi$_2$ or SrMnBi$_2$, in which the ZFC-FC splitting occurs when the field parallel to c axis. This difference implies that the magnetic structure of BaMnBi$_2$ may be slightly different from that of CaMnBi$_2$ or SrMnBi$_2$.

\indent The carriers of a metal in magnetic field are subject to the Lorentz force. The Lorentz force affects the carriers momentum components in the plane perpendicular to the field and MR is determined by the mobility in this plane. The carriers in a quasi-two-dimensional material will only be affected by the magnetic field component $B|cos\theta|$. Fig.3(a) shows the angular-dependent in-plane MR measured in different magnetic fields. When $\theta=0^0,180^0$, that is magnetic field parallel to c axis, MR has the maximum value. With the increase in the polar angle $\theta$, the MR decreases gradually and reaches the minimum value when $\theta=90^0$. The cosine function behavior of the whole curve indicates the existence of an anisotropic quasi-two-dimensional Fermi surface in BaMnBi$_2$. The small deviation implies the existence of 3D electronic transport in BaMnBi$_2$.

\begin{figure}[b]
\centering
\includegraphics[width=0.48\textwidth]{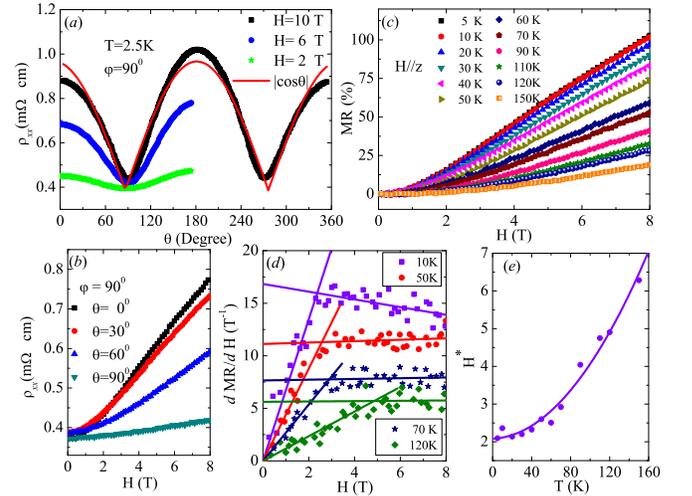}
\caption{(Color online) (a) Polar angle $\theta$ dependence of in-plane resistivity $\rho_{xx}$ at H=10,6,2T. The temperature and azimuthal angle are fixed at T=2.5K and $\varphi=90^0$ respectively. The red curve is the fitting of cosine function. (b) Magnetic field H dependence of in-plane resistivity $\rho_{xx}$ with different polar angle $\theta$ at 2.5K. (c) In-plane magnetoresistance MR vs magnetic field H with the temperature from 5K to 150K. (d) The field derivative of the MR as a function of magnetic field H. The straight lines in low field regions are linear fitting results using the relationship $MR=A_2H^2$. Lines in high field regions are fitting results using $MR=A_1H+O(H^2)$. The intersection of fitting lines in low field and high field is defined as critical point H$^*$. (e) Temperature dependence of critical point H$^*$. The solid curve is the fit to the equation $H^*=(E_F+k_BT)^2/2ev_{F}^2\hbar$.}
\end{figure}

\indent Fig.3(c) shows the in-plane MR as a function of magnetic field at different temperatures. The MR exhibits quantum linear behavior in high field and semiclassical quadratic dependence in low field\cite{abrikosov1988fundamentals}. The transition from semiclassical to quantum can be seen more clearly from the field derivative of the MR [Fig.3(d)]. Initially, $dMR/dH$ is proportional to H, which indicates a semiclassical $H^2$-dependent MR.  With the increase of magnetic field, the field will crosses a critical point $H^*$ and $dMR/dH$ nearly becomes a constant. This implies that MR follows linear field dependence plus a very small quadratic term in high field, namely $MR=A_1H+O(H^2)$. The linear MR suggests the existence of Dirac fermions in BaMnBi$_2$\cite{PhysRevB.58.2788,abrikosov2000quantum,hu2008classical,PhysRevLett.88.066602}.

\indent A sufficiently strong perpendicular magnetic field can cause the complete quantization of orbit of Dirac fermions and the quantized Landau level can be described as $E_n=sgn(n)v_F\sqrt{2e\hbar H|n|}$, where $v_F$ is the Fermi velocity and $n$ is the Landau index. So the energy splitting between the first and lowest Landau level is $\Delta=\pm v_F\sqrt{2e\hbar H}$. At a specific temperature T, with the increase of field H, the energy splitting becomes larger than the Fermi energy $E_F$ and the thermal fluctuations $k_BT$. As a result, all carriers will occupy the lowest Landau level and lead to the quantum linear MR. Combined these two relationships, we can get the temperature dependence of the critical magnetic field is $H^*=(E_F+k_BT)^2/2e\hbar v_F^2$\cite{PhysRevLett.106.217004}. Fig.3(e) presents the relationship between T and $H^*$ and can be well fitted by this equation. This also supports the presence of Dirac fermions in BaMnBi$_2$.

\section{Summary}
\indent In summary, single crystals of BaMnBi$_2$ have been grown successfully. Resistivity, Hall resistivity, magnetic property and magnetoresistance have been measured and analyzed. Compared to CaMnBi$_2$ and SrMnBi$_2$, the different behavior of magnetization brings the possible different magnetic structure of BaMnBi$_2$. Quasi-2D electronic transport is observed in angle-dependent MR. The crossover from semiclassical parabolic field-dependent MR in low field to linear field-dependent MR in high field can be explained by combining semiclassical and quantum magnetoresistance. Our results indicate the existence of anisotropic Dirac fermions in BaMnBi$_2$.

\emph{Note added.} When the paper is being finalized, we notice one similar related work on BaMnBi$_2$\cite{Petrovic2016}, where the Shubnikov-de Hass oscillation is also observed under high magnetic field.

\section{Acknowledgments}
\indent We thank professor J.F. Wang for his help in our attempt to obtain high field data. This work is supported by the National Natural Science Foundation of China (No.11574391), the Fundamental Research Funds for the Central Universities, and the Research Funds of Renmin University of China (No. 14XNLQ07).
\bibliography{bibtex}

\begin{thebibliography}{48}
\expandafter\ifx\csname natexlab\endcsname\relax\def\natexlab#1{#1}\fi
\expandafter\ifx\csname bibnamefont\endcsname\relax
  \def\bibnamefont#1{#1}\fi
\expandafter\ifx\csname bibfnamefont\endcsname\relax
  \def\bibfnamefont#1{#1}\fi
\expandafter\ifx\csname citenamefont\endcsname\relax
  \def\citenamefont#1{#1}\fi
\expandafter\ifx\csname url\endcsname\relax
  \def\url#1{\texttt{#1}}\fi
\expandafter\ifx\csname urlprefix\endcsname\relax\def\urlprefix{URL }\fi
\providecommand{\bibinfo}[2]{#2}
\providecommand{\eprint}[2][]{\url{#2}}

\bibitem[{\citenamefont{Wehling et~al.}(2014)\citenamefont{Wehling,
  Black-Schaffer, and Balatsky}}]{wehling2014dirac}
\bibinfo{author}{\bibfnamefont{T.}~\bibnamefont{Wehling}},
  \bibinfo{author}{\bibfnamefont{A.~M.} \bibnamefont{Black-Schaffer}},
  \bibnamefont{and} \bibinfo{author}{\bibfnamefont{A.~V.}
  \bibnamefont{Balatsky}}, \bibinfo{journal}{Adv. Phys.}
  \textbf{\bibinfo{volume}{63}}, \bibinfo{pages}{1} (\bibinfo{year}{2014}).

\bibitem[{\citenamefont{Castro~Neto et~al.}(2009)\citenamefont{Castro~Neto,
  Guinea, Peres, Novoselov, and Geim}}]{RevModPhys.81.109}
\bibinfo{author}{\bibfnamefont{A.~H.} \bibnamefont{Castro~Neto}},
  \bibinfo{author}{\bibfnamefont{F.}~\bibnamefont{Guinea}},
  \bibinfo{author}{\bibfnamefont{N.~M.~R.} \bibnamefont{Peres}},
  \bibinfo{author}{\bibfnamefont{K.~S.} \bibnamefont{Novoselov}},
  \bibnamefont{and} \bibinfo{author}{\bibfnamefont{A.~K.} \bibnamefont{Geim}},
  \bibinfo{journal}{Rev. Mod. Phys.} \textbf{\bibinfo{volume}{81}},
  \bibinfo{pages}{109} (\bibinfo{year}{2009}).

\bibitem[{\citenamefont{Hasan and Kane}(2010)}]{RevModPhys.82.3045}
\bibinfo{author}{\bibfnamefont{M.~Z.} \bibnamefont{Hasan}} \bibnamefont{and}
  \bibinfo{author}{\bibfnamefont{C.~L.} \bibnamefont{Kane}},
  \bibinfo{journal}{Rev. Mod. Phys.} \textbf{\bibinfo{volume}{82}},
  \bibinfo{pages}{3045} (\bibinfo{year}{2010}).

\bibitem[{\citenamefont{Qi and Zhang}(2011)}]{RevModPhys.83.1057}
\bibinfo{author}{\bibfnamefont{X.-L.} \bibnamefont{Qi}} \bibnamefont{and}
  \bibinfo{author}{\bibfnamefont{S.-C.} \bibnamefont{Zhang}},
  \bibinfo{journal}{Rev. Mod. Phys.} \textbf{\bibinfo{volume}{83}},
  \bibinfo{pages}{1057} (\bibinfo{year}{2011}).

\bibitem[{\citenamefont{Tsuei and Kirtley}(2000)}]{RevModPhys.72.969}
\bibinfo{author}{\bibfnamefont{C.~C.} \bibnamefont{Tsuei}} \bibnamefont{and}
  \bibinfo{author}{\bibfnamefont{J.~R.} \bibnamefont{Kirtley}},
  \bibinfo{journal}{Rev. Mod. Phys.} \textbf{\bibinfo{volume}{72}},
  \bibinfo{pages}{969} (\bibinfo{year}{2000}).

\bibitem[{\citenamefont{Liu et~al.}(2014)\citenamefont{Liu, Zhou, Zhang, Wang,
  Weng, Prabhakaran, Mo, Shen, Fang, Dai et~al.}}]{liu2014discovery}
\bibinfo{author}{\bibfnamefont{Z.}~\bibnamefont{Liu}},
  \bibinfo{author}{\bibfnamefont{B.}~\bibnamefont{Zhou}},
  \bibinfo{author}{\bibfnamefont{Y.}~\bibnamefont{Zhang}},
  \bibinfo{author}{\bibfnamefont{Z.}~\bibnamefont{Wang}},
  \bibinfo{author}{\bibfnamefont{H.}~\bibnamefont{Weng}},
  \bibinfo{author}{\bibfnamefont{D.}~\bibnamefont{Prabhakaran}},
  \bibinfo{author}{\bibfnamefont{S.-K.} \bibnamefont{Mo}},
  \bibinfo{author}{\bibfnamefont{Z.}~\bibnamefont{Shen}},
  \bibinfo{author}{\bibfnamefont{Z.}~\bibnamefont{Fang}},
  \bibinfo{author}{\bibfnamefont{X.}~\bibnamefont{Dai}}, \bibnamefont{et~al.},
  \bibinfo{journal}{Science} \textbf{\bibinfo{volume}{343}},
  \bibinfo{pages}{864} (\bibinfo{year}{2014}).

\bibitem[{\citenamefont{Borisenko et~al.}(2014)\citenamefont{Borisenko, Gibson,
  Evtushinsky, Zabolotnyy, B\"uchner, and Cava}}]{PhysRevLett.113.027603}
\bibinfo{author}{\bibfnamefont{S.}~\bibnamefont{Borisenko}},
  \bibinfo{author}{\bibfnamefont{Q.}~\bibnamefont{Gibson}},
  \bibinfo{author}{\bibfnamefont{D.}~\bibnamefont{Evtushinsky}},
  \bibinfo{author}{\bibfnamefont{V.}~\bibnamefont{Zabolotnyy}},
  \bibinfo{author}{\bibfnamefont{B.}~\bibnamefont{B\"uchner}},
  \bibnamefont{and} \bibinfo{author}{\bibfnamefont{R.~J.} \bibnamefont{Cava}},
  \bibinfo{journal}{Phys. Rev. Lett.} \textbf{\bibinfo{volume}{113}},
  \bibinfo{pages}{027603} (\bibinfo{year}{2014}).

\bibitem[{\citenamefont{Neupane et~al.}(2014)\citenamefont{Neupane, Xu, Sankar,
  Alidoust, Bian, Liu, Belopolski, Chang, Jeng, Lin
  et~al.}}]{neupane2014observation}
\bibinfo{author}{\bibfnamefont{M.}~\bibnamefont{Neupane}},
  \bibinfo{author}{\bibfnamefont{S.-Y.} \bibnamefont{Xu}},
  \bibinfo{author}{\bibfnamefont{R.}~\bibnamefont{Sankar}},
  \bibinfo{author}{\bibfnamefont{N.}~\bibnamefont{Alidoust}},
  \bibinfo{author}{\bibfnamefont{G.}~\bibnamefont{Bian}},
  \bibinfo{author}{\bibfnamefont{C.}~\bibnamefont{Liu}},
  \bibinfo{author}{\bibfnamefont{I.}~\bibnamefont{Belopolski}},
  \bibinfo{author}{\bibfnamefont{T.-R.} \bibnamefont{Chang}},
  \bibinfo{author}{\bibfnamefont{H.-T.} \bibnamefont{Jeng}},
  \bibinfo{author}{\bibfnamefont{H.}~\bibnamefont{Lin}}, \bibnamefont{et~al.},
  \bibinfo{journal}{Nat. Commun.} \textbf{\bibinfo{volume}{5}},
  \bibinfo{pages}{3786} (\bibinfo{year}{2014}).

\bibitem[{\citenamefont{Liang et~al.}(2015)\citenamefont{Liang, Gibson, Ali,
  Liu, Cava, and Ong}}]{liang2015ultrahigh}
\bibinfo{author}{\bibfnamefont{T.}~\bibnamefont{Liang}},
  \bibinfo{author}{\bibfnamefont{Q.}~\bibnamefont{Gibson}},
  \bibinfo{author}{\bibfnamefont{M.~N.} \bibnamefont{Ali}},
  \bibinfo{author}{\bibfnamefont{M.}~\bibnamefont{Liu}},
  \bibinfo{author}{\bibfnamefont{R.}~\bibnamefont{Cava}}, \bibnamefont{and}
  \bibinfo{author}{\bibfnamefont{N.}~\bibnamefont{Ong}}, \bibinfo{journal}{Nat.
  Mater.} \textbf{\bibinfo{volume}{14}}, \bibinfo{pages}{280}
  (\bibinfo{year}{2015}).

\bibitem[{\citenamefont{He et~al.}(2014)\citenamefont{He, Hong, Dong, Pan,
  Zhang, Zhang, and Li}}]{PhysRevLett.113.246402}
\bibinfo{author}{\bibfnamefont{L.~P.} \bibnamefont{He}},
  \bibinfo{author}{\bibfnamefont{X.~C.} \bibnamefont{Hong}},
  \bibinfo{author}{\bibfnamefont{J.~K.} \bibnamefont{Dong}},
  \bibinfo{author}{\bibfnamefont{J.}~\bibnamefont{Pan}},
  \bibinfo{author}{\bibfnamefont{Z.}~\bibnamefont{Zhang}},
  \bibinfo{author}{\bibfnamefont{J.}~\bibnamefont{Zhang}}, \bibnamefont{and}
  \bibinfo{author}{\bibfnamefont{S.~Y.} \bibnamefont{Li}},
  \bibinfo{journal}{Phys. Rev. Lett.} \textbf{\bibinfo{volume}{113}},
  \bibinfo{pages}{246402} (\bibinfo{year}{2014}).

\bibitem[{\citenamefont{Wan et~al.}(2011)\citenamefont{Wan, Turner, Vishwanath,
  and Savrasov}}]{PhysRevB.83.205101}
\bibinfo{author}{\bibfnamefont{X.}~\bibnamefont{Wan}},
  \bibinfo{author}{\bibfnamefont{A.~M.} \bibnamefont{Turner}},
  \bibinfo{author}{\bibfnamefont{A.}~\bibnamefont{Vishwanath}},
  \bibnamefont{and} \bibinfo{author}{\bibfnamefont{S.~Y.}
  \bibnamefont{Savrasov}}, \bibinfo{journal}{Phys. Rev. B}
  \textbf{\bibinfo{volume}{83}}, \bibinfo{pages}{205101}
  (\bibinfo{year}{2011}).

\bibitem[{\citenamefont{Lv et~al.}(2015{\natexlab{a}})\citenamefont{Lv, Weng,
  Fu, Wang, Miao, Ma, Richard, Huang, Zhao, Chen et~al.}}]{PhysRevX.5.031013}
\bibinfo{author}{\bibfnamefont{B.~Q.} \bibnamefont{Lv}},
  \bibinfo{author}{\bibfnamefont{H.~M.} \bibnamefont{Weng}},
  \bibinfo{author}{\bibfnamefont{B.~B.} \bibnamefont{Fu}},
  \bibinfo{author}{\bibfnamefont{X.~P.} \bibnamefont{Wang}},
  \bibinfo{author}{\bibfnamefont{H.}~\bibnamefont{Miao}},
  \bibinfo{author}{\bibfnamefont{J.}~\bibnamefont{Ma}},
  \bibinfo{author}{\bibfnamefont{P.}~\bibnamefont{Richard}},
  \bibinfo{author}{\bibfnamefont{X.~C.} \bibnamefont{Huang}},
  \bibinfo{author}{\bibfnamefont{L.~X.} \bibnamefont{Zhao}},
  \bibinfo{author}{\bibfnamefont{G.~F.} \bibnamefont{Chen}},
  \bibnamefont{et~al.}, \bibinfo{journal}{Phys. Rev. X}
  \textbf{\bibinfo{volume}{5}}, \bibinfo{pages}{031013}
  (\bibinfo{year}{2015}{\natexlab{a}}).

\bibitem[{\citenamefont{Xu et~al.}(2015{\natexlab{a}})\citenamefont{Xu,
  Belopolski, Alidoust, Neupane, Bian, Zhang, Sankar, Chang, Yuan, Lee
  et~al.}}]{xu2015discovery}
\bibinfo{author}{\bibfnamefont{S.-Y.} \bibnamefont{Xu}},
  \bibinfo{author}{\bibfnamefont{I.}~\bibnamefont{Belopolski}},
  \bibinfo{author}{\bibfnamefont{N.}~\bibnamefont{Alidoust}},
  \bibinfo{author}{\bibfnamefont{M.}~\bibnamefont{Neupane}},
  \bibinfo{author}{\bibfnamefont{G.}~\bibnamefont{Bian}},
  \bibinfo{author}{\bibfnamefont{C.}~\bibnamefont{Zhang}},
  \bibinfo{author}{\bibfnamefont{R.}~\bibnamefont{Sankar}},
  \bibinfo{author}{\bibfnamefont{G.}~\bibnamefont{Chang}},
  \bibinfo{author}{\bibfnamefont{Z.}~\bibnamefont{Yuan}},
  \bibinfo{author}{\bibfnamefont{C.-C.} \bibnamefont{Lee}},
  \bibnamefont{et~al.}, \bibinfo{journal}{Science}
  \textbf{\bibinfo{volume}{349}}, \bibinfo{pages}{613}
  (\bibinfo{year}{2015}{\natexlab{a}}).

\bibitem[{\citenamefont{Lv et~al.}(2015{\natexlab{b}})\citenamefont{Lv, Xu,
  Weng, Ma, Richard, Huang, Zhao, Chen, Matt, Bisti
  et~al.}}]{lv2015observation}
\bibinfo{author}{\bibfnamefont{B.}~\bibnamefont{Lv}},
  \bibinfo{author}{\bibfnamefont{N.}~\bibnamefont{Xu}},
  \bibinfo{author}{\bibfnamefont{H.}~\bibnamefont{Weng}},
  \bibinfo{author}{\bibfnamefont{J.}~\bibnamefont{Ma}},
  \bibinfo{author}{\bibfnamefont{P.}~\bibnamefont{Richard}},
  \bibinfo{author}{\bibfnamefont{X.}~\bibnamefont{Huang}},
  \bibinfo{author}{\bibfnamefont{L.}~\bibnamefont{Zhao}},
  \bibinfo{author}{\bibfnamefont{G.}~\bibnamefont{Chen}},
  \bibinfo{author}{\bibfnamefont{C.}~\bibnamefont{Matt}},
  \bibinfo{author}{\bibfnamefont{F.}~\bibnamefont{Bisti}},
  \bibnamefont{et~al.}, \bibinfo{journal}{Nat. Phys.}
  \textbf{\bibinfo{volume}{11}}, \bibinfo{pages}{724}
  (\bibinfo{year}{2015}{\natexlab{b}}).

\bibitem[{\citenamefont{Huang et~al.}(2015{\natexlab{a}})\citenamefont{Huang,
  Zhao, Long, Wang, Chen, Yang, Liang, Xue, Weng, Fang
  et~al.}}]{PhysRevX.5.031023}
\bibinfo{author}{\bibfnamefont{X.}~\bibnamefont{Huang}},
  \bibinfo{author}{\bibfnamefont{L.}~\bibnamefont{Zhao}},
  \bibinfo{author}{\bibfnamefont{Y.}~\bibnamefont{Long}},
  \bibinfo{author}{\bibfnamefont{P.}~\bibnamefont{Wang}},
  \bibinfo{author}{\bibfnamefont{D.}~\bibnamefont{Chen}},
  \bibinfo{author}{\bibfnamefont{Z.}~\bibnamefont{Yang}},
  \bibinfo{author}{\bibfnamefont{H.}~\bibnamefont{Liang}},
  \bibinfo{author}{\bibfnamefont{M.}~\bibnamefont{Xue}},
  \bibinfo{author}{\bibfnamefont{H.}~\bibnamefont{Weng}},
  \bibinfo{author}{\bibfnamefont{Z.}~\bibnamefont{Fang}}, \bibnamefont{et~al.},
  \bibinfo{journal}{Phys. Rev. X} \textbf{\bibinfo{volume}{5}},
  \bibinfo{pages}{031023} (\bibinfo{year}{2015}{\natexlab{a}}).

\bibitem[{\citenamefont{Yang et~al.}(2015{\natexlab{a}})\citenamefont{Yang,
  Liu, Sun, Peng, Yang, Zhang, Zhou, Zhang, Guo, Rahn et~al.}}]{yang2015weyl}
\bibinfo{author}{\bibfnamefont{L.}~\bibnamefont{Yang}},
  \bibinfo{author}{\bibfnamefont{Z.}~\bibnamefont{Liu}},
  \bibinfo{author}{\bibfnamefont{Y.}~\bibnamefont{Sun}},
  \bibinfo{author}{\bibfnamefont{H.}~\bibnamefont{Peng}},
  \bibinfo{author}{\bibfnamefont{H.}~\bibnamefont{Yang}},
  \bibinfo{author}{\bibfnamefont{T.}~\bibnamefont{Zhang}},
  \bibinfo{author}{\bibfnamefont{B.}~\bibnamefont{Zhou}},
  \bibinfo{author}{\bibfnamefont{Y.}~\bibnamefont{Zhang}},
  \bibinfo{author}{\bibfnamefont{Y.}~\bibnamefont{Guo}},
  \bibinfo{author}{\bibfnamefont{M.}~\bibnamefont{Rahn}}, \bibnamefont{et~al.},
  \bibinfo{journal}{Nat. Phys.} \textbf{\bibinfo{volume}{11}},
  \bibinfo{pages}{728} (\bibinfo{year}{2015}{\natexlab{a}}).

\bibitem[{\citenamefont{Huang et~al.}(2015{\natexlab{b}})\citenamefont{Huang,
  Xu, Belopolski, Lee, Chang, Wang, Alidoust, Bian, Neupane, Zhang
  et~al.}}]{huang2015weyl}
\bibinfo{author}{\bibfnamefont{S.-M.} \bibnamefont{Huang}},
  \bibinfo{author}{\bibfnamefont{S.-Y.} \bibnamefont{Xu}},
  \bibinfo{author}{\bibfnamefont{I.}~\bibnamefont{Belopolski}},
  \bibinfo{author}{\bibfnamefont{C.-C.} \bibnamefont{Lee}},
  \bibinfo{author}{\bibfnamefont{G.}~\bibnamefont{Chang}},
  \bibinfo{author}{\bibfnamefont{B.}~\bibnamefont{Wang}},
  \bibinfo{author}{\bibfnamefont{N.}~\bibnamefont{Alidoust}},
  \bibinfo{author}{\bibfnamefont{G.}~\bibnamefont{Bian}},
  \bibinfo{author}{\bibfnamefont{M.}~\bibnamefont{Neupane}},
  \bibinfo{author}{\bibfnamefont{C.}~\bibnamefont{Zhang}},
  \bibnamefont{et~al.}, \bibinfo{journal}{Nat. Commun.}
  \textbf{\bibinfo{volume}{6}}, \bibinfo{pages}{7373}
  (\bibinfo{year}{2015}{\natexlab{b}}).

\bibitem[{\citenamefont{Xu et~al.}(2015{\natexlab{b}})\citenamefont{Xu,
  Belopolski, Sanchez, Zhang, Chang, Guo, Bian, Yuan, Lu, Chang
  et~al.}}]{xu2015experimental}
\bibinfo{author}{\bibfnamefont{S.-Y.} \bibnamefont{Xu}},
  \bibinfo{author}{\bibfnamefont{I.}~\bibnamefont{Belopolski}},
  \bibinfo{author}{\bibfnamefont{D.~S.} \bibnamefont{Sanchez}},
  \bibinfo{author}{\bibfnamefont{C.}~\bibnamefont{Zhang}},
  \bibinfo{author}{\bibfnamefont{G.}~\bibnamefont{Chang}},
  \bibinfo{author}{\bibfnamefont{C.}~\bibnamefont{Guo}},
  \bibinfo{author}{\bibfnamefont{G.}~\bibnamefont{Bian}},
  \bibinfo{author}{\bibfnamefont{Z.}~\bibnamefont{Yuan}},
  \bibinfo{author}{\bibfnamefont{H.}~\bibnamefont{Lu}},
  \bibinfo{author}{\bibfnamefont{T.-R.} \bibnamefont{Chang}},
  \bibnamefont{et~al.}, \bibinfo{journal}{Sci. Adv.}
  \textbf{\bibinfo{volume}{1}}, \bibinfo{pages}{e1501092}
  (\bibinfo{year}{2015}{\natexlab{b}}).

\bibitem[{\citenamefont{Weng et~al.}(2015)\citenamefont{Weng, Fang, Fang,
  Bernevig, and Dai}}]{PhysRevX.5.011029}
\bibinfo{author}{\bibfnamefont{H.}~\bibnamefont{Weng}},
  \bibinfo{author}{\bibfnamefont{C.}~\bibnamefont{Fang}},
  \bibinfo{author}{\bibfnamefont{Z.}~\bibnamefont{Fang}},
  \bibinfo{author}{\bibfnamefont{B.~A.} \bibnamefont{Bernevig}},
  \bibnamefont{and} \bibinfo{author}{\bibfnamefont{X.}~\bibnamefont{Dai}},
  \bibinfo{journal}{Phys. Rev. X} \textbf{\bibinfo{volume}{5}},
  \bibinfo{pages}{011029} (\bibinfo{year}{2015}).

\bibitem[{\citenamefont{Liu et~al.}(2016)\citenamefont{Liu, Yang, Sun, Zhang,
  Peng, Yang, Chen, Zhang, Guo, Prabhakaran et~al.}}]{liu2016evolution}
\bibinfo{author}{\bibfnamefont{Z.}~\bibnamefont{Liu}},
  \bibinfo{author}{\bibfnamefont{L.}~\bibnamefont{Yang}},
  \bibinfo{author}{\bibfnamefont{Y.}~\bibnamefont{Sun}},
  \bibinfo{author}{\bibfnamefont{T.}~\bibnamefont{Zhang}},
  \bibinfo{author}{\bibfnamefont{H.}~\bibnamefont{Peng}},
  \bibinfo{author}{\bibfnamefont{H.}~\bibnamefont{Yang}},
  \bibinfo{author}{\bibfnamefont{C.}~\bibnamefont{Chen}},
  \bibinfo{author}{\bibfnamefont{Y.}~\bibnamefont{Zhang}},
  \bibinfo{author}{\bibfnamefont{Y.}~\bibnamefont{Guo}},
  \bibinfo{author}{\bibfnamefont{D.}~\bibnamefont{Prabhakaran}},
  \bibnamefont{et~al.}, \bibinfo{journal}{Nat. Mater.}
  \textbf{\bibinfo{volume}{15}}, \bibinfo{pages}{27} (\bibinfo{year}{2016}).

\bibitem[{\citenamefont{Shekhar
  et~al.}(2015{\natexlab{a}})\citenamefont{Shekhar, Nayak, Sun, Schmidt,
  Nicklas, Leermakers, Zeitler, Skourski, Wosnitza, Liu
  et~al.}}]{shekhar2015extremely}
\bibinfo{author}{\bibfnamefont{C.}~\bibnamefont{Shekhar}},
  \bibinfo{author}{\bibfnamefont{A.~K.} \bibnamefont{Nayak}},
  \bibinfo{author}{\bibfnamefont{Y.}~\bibnamefont{Sun}},
  \bibinfo{author}{\bibfnamefont{M.}~\bibnamefont{Schmidt}},
  \bibinfo{author}{\bibfnamefont{M.}~\bibnamefont{Nicklas}},
  \bibinfo{author}{\bibfnamefont{I.}~\bibnamefont{Leermakers}},
  \bibinfo{author}{\bibfnamefont{U.}~\bibnamefont{Zeitler}},
  \bibinfo{author}{\bibfnamefont{Y.}~\bibnamefont{Skourski}},
  \bibinfo{author}{\bibfnamefont{J.}~\bibnamefont{Wosnitza}},
  \bibinfo{author}{\bibfnamefont{Z.}~\bibnamefont{Liu}}, \bibnamefont{et~al.},
  \bibinfo{journal}{Nat. Phys.} \textbf{\bibinfo{volume}{11}},
  \bibinfo{pages}{645} (\bibinfo{year}{2015}{\natexlab{a}}).

\bibitem[{\citenamefont{Shekhar
  et~al.}(2015{\natexlab{b}})\citenamefont{Shekhar, Arnold, Wu, Sun, Schmidt,
  Kumar, Grushin, Bardarson, Reis, Naumann et~al.}}]{shekhar2015large}
\bibinfo{author}{\bibfnamefont{C.}~\bibnamefont{Shekhar}},
  \bibinfo{author}{\bibfnamefont{F.}~\bibnamefont{Arnold}},
  \bibinfo{author}{\bibfnamefont{S.-C.} \bibnamefont{Wu}},
  \bibinfo{author}{\bibfnamefont{Y.}~\bibnamefont{Sun}},
  \bibinfo{author}{\bibfnamefont{M.}~\bibnamefont{Schmidt}},
  \bibinfo{author}{\bibfnamefont{N.}~\bibnamefont{Kumar}},
  \bibinfo{author}{\bibfnamefont{A.~G.} \bibnamefont{Grushin}},
  \bibinfo{author}{\bibfnamefont{J.~H.} \bibnamefont{Bardarson}},
  \bibinfo{author}{\bibfnamefont{R.~D.~d.} \bibnamefont{Reis}},
  \bibinfo{author}{\bibfnamefont{M.}~\bibnamefont{Naumann}},
  \bibnamefont{et~al.}, \bibinfo{journal}{arXiv preprint arXiv:1506.06577}
  (\bibinfo{year}{2015}{\natexlab{b}}).

\bibitem[{\citenamefont{Yang et~al.}(2015{\natexlab{b}})\citenamefont{Yang,
  Liu, Wang, Zheng, and Xu}}]{yang2015chiral}
\bibinfo{author}{\bibfnamefont{X.}~\bibnamefont{Yang}},
  \bibinfo{author}{\bibfnamefont{Y.}~\bibnamefont{Liu}},
  \bibinfo{author}{\bibfnamefont{Z.}~\bibnamefont{Wang}},
  \bibinfo{author}{\bibfnamefont{Y.}~\bibnamefont{Zheng}}, \bibnamefont{and}
  \bibinfo{author}{\bibfnamefont{Z.-a.} \bibnamefont{Xu}},
  \bibinfo{journal}{arXiv preprint arXiv:1506.03190}
  (\bibinfo{year}{2015}{\natexlab{b}}).

\bibitem[{\citenamefont{Sun et~al.}(2015)\citenamefont{Sun, Wu, and
  Yan}}]{PhysRevB.92.115428}
\bibinfo{author}{\bibfnamefont{Y.}~\bibnamefont{Sun}},
  \bibinfo{author}{\bibfnamefont{S.-C.} \bibnamefont{Wu}}, \bibnamefont{and}
  \bibinfo{author}{\bibfnamefont{B.}~\bibnamefont{Yan}},
  \bibinfo{journal}{Phys. Rev. B} \textbf{\bibinfo{volume}{92}},
  \bibinfo{pages}{115428} (\bibinfo{year}{2015}).

\bibitem[{\citenamefont{Zhang et~al.}(2005)\citenamefont{Zhang, Tan, Stormer,
  and Kim}}]{zhang2005experimental}
\bibinfo{author}{\bibfnamefont{Y.}~\bibnamefont{Zhang}},
  \bibinfo{author}{\bibfnamefont{Y.-W.} \bibnamefont{Tan}},
  \bibinfo{author}{\bibfnamefont{H.~L.} \bibnamefont{Stormer}},
  \bibnamefont{and} \bibinfo{author}{\bibfnamefont{P.}~\bibnamefont{Kim}},
  \bibinfo{journal}{Nature} \textbf{\bibinfo{volume}{438}},
  \bibinfo{pages}{201} (\bibinfo{year}{2005}).

\bibitem[{\citenamefont{Miller et~al.}(2009)\citenamefont{Miller, Kubista,
  Rutter, Ruan, de~Heer, First, and Stroscio}}]{miller2009observing}
\bibinfo{author}{\bibfnamefont{D.~L.} \bibnamefont{Miller}},
  \bibinfo{author}{\bibfnamefont{K.~D.} \bibnamefont{Kubista}},
  \bibinfo{author}{\bibfnamefont{G.~M.} \bibnamefont{Rutter}},
  \bibinfo{author}{\bibfnamefont{M.}~\bibnamefont{Ruan}},
  \bibinfo{author}{\bibfnamefont{W.~A.} \bibnamefont{de~Heer}},
  \bibinfo{author}{\bibfnamefont{P.~N.} \bibnamefont{First}}, \bibnamefont{and}
  \bibinfo{author}{\bibfnamefont{J.~A.} \bibnamefont{Stroscio}},
  \bibinfo{journal}{Science} \textbf{\bibinfo{volume}{324}},
  \bibinfo{pages}{924} (\bibinfo{year}{2009}).

\bibitem[{\citenamefont{Wang et~al.}(2012)\citenamefont{Wang, Graf, Wang, Lei,
  Tozer, and Petrovic}}]{PhysRevB.85.041101}
\bibinfo{author}{\bibfnamefont{K.}~\bibnamefont{Wang}},
  \bibinfo{author}{\bibfnamefont{D.}~\bibnamefont{Graf}},
  \bibinfo{author}{\bibfnamefont{L.}~\bibnamefont{Wang}},
  \bibinfo{author}{\bibfnamefont{H.}~\bibnamefont{Lei}},
  \bibinfo{author}{\bibfnamefont{S.~W.} \bibnamefont{Tozer}}, \bibnamefont{and}
  \bibinfo{author}{\bibfnamefont{C.}~\bibnamefont{Petrovic}},
  \bibinfo{journal}{Phys. Rev. B} \textbf{\bibinfo{volume}{85}},
  \bibinfo{pages}{041101} (\bibinfo{year}{2012}).

\bibitem[{\citenamefont{He et~al.}(2012)\citenamefont{He, Wang, and
  Chen}}]{he2012giant}
\bibinfo{author}{\bibfnamefont{J.}~\bibnamefont{He}},
  \bibinfo{author}{\bibfnamefont{D.}~\bibnamefont{Wang}}, \bibnamefont{and}
  \bibinfo{author}{\bibfnamefont{G.}~\bibnamefont{Chen}},
  \bibinfo{journal}{Appl. Phys. Lett.} \textbf{\bibinfo{volume}{100}},
  \bibinfo{pages}{112405} (\bibinfo{year}{2012}).

\bibitem[{\citenamefont{Feng et~al.}(2014)\citenamefont{Feng, Wang, Chen, Shi,
  Xie, Yi, Liang, He, He, Peng et~al.}}]{feng2014strong}
\bibinfo{author}{\bibfnamefont{Y.}~\bibnamefont{Feng}},
  \bibinfo{author}{\bibfnamefont{Z.}~\bibnamefont{Wang}},
  \bibinfo{author}{\bibfnamefont{C.}~\bibnamefont{Chen}},
  \bibinfo{author}{\bibfnamefont{Y.}~\bibnamefont{Shi}},
  \bibinfo{author}{\bibfnamefont{Z.}~\bibnamefont{Xie}},
  \bibinfo{author}{\bibfnamefont{H.}~\bibnamefont{Yi}},
  \bibinfo{author}{\bibfnamefont{A.}~\bibnamefont{Liang}},
  \bibinfo{author}{\bibfnamefont{S.}~\bibnamefont{He}},
  \bibinfo{author}{\bibfnamefont{J.}~\bibnamefont{He}},
  \bibinfo{author}{\bibfnamefont{Y.}~\bibnamefont{Peng}}, \bibnamefont{et~al.},
  \bibinfo{journal}{Sci. Rep.} \textbf{\bibinfo{volume}{4}},
  \bibinfo{pages}{5385} (\bibinfo{year}{2014}).

\bibitem[{\citenamefont{Lee et~al.}(2013)\citenamefont{Lee, Farhan, Kim, and
  Shim}}]{PhysRevB.87.245104}
\bibinfo{author}{\bibfnamefont{G.}~\bibnamefont{Lee}},
  \bibinfo{author}{\bibfnamefont{M.~A.} \bibnamefont{Farhan}},
  \bibinfo{author}{\bibfnamefont{J.~S.} \bibnamefont{Kim}}, \bibnamefont{and}
  \bibinfo{author}{\bibfnamefont{J.~H.} \bibnamefont{Shim}},
  \bibinfo{journal}{Phys. Rev. B} \textbf{\bibinfo{volume}{87}},
  \bibinfo{pages}{245104} (\bibinfo{year}{2013}).

\bibitem[{\citenamefont{Park et~al.}(2011)\citenamefont{Park, Lee,
  Wolff-Fabris, Koh, Eom, Kim, Farhan, Jo, Kim, Shim
  et~al.}}]{PhysRevLett.107.126402}
\bibinfo{author}{\bibfnamefont{J.}~\bibnamefont{Park}},
  \bibinfo{author}{\bibfnamefont{G.}~\bibnamefont{Lee}},
  \bibinfo{author}{\bibfnamefont{F.}~\bibnamefont{Wolff-Fabris}},
  \bibinfo{author}{\bibfnamefont{Y.~Y.} \bibnamefont{Koh}},
  \bibinfo{author}{\bibfnamefont{M.~J.} \bibnamefont{Eom}},
  \bibinfo{author}{\bibfnamefont{Y.~K.} \bibnamefont{Kim}},
  \bibinfo{author}{\bibfnamefont{M.~A.} \bibnamefont{Farhan}},
  \bibinfo{author}{\bibfnamefont{Y.~J.} \bibnamefont{Jo}},
  \bibinfo{author}{\bibfnamefont{C.}~\bibnamefont{Kim}},
  \bibinfo{author}{\bibfnamefont{J.~H.} \bibnamefont{Shim}},
  \bibnamefont{et~al.}, \bibinfo{journal}{Phys. Rev. Lett.}
  \textbf{\bibinfo{volume}{107}}, \bibinfo{pages}{126402}
  (\bibinfo{year}{2011}).

\bibitem[{\citenamefont{Wang et~al.}(2011{\natexlab{a}})\citenamefont{Wang,
  Graf, Lei, Tozer, and Petrovic}}]{PhysRevB.84.220401}
\bibinfo{author}{\bibfnamefont{K.}~\bibnamefont{Wang}},
  \bibinfo{author}{\bibfnamefont{D.}~\bibnamefont{Graf}},
  \bibinfo{author}{\bibfnamefont{H.}~\bibnamefont{Lei}},
  \bibinfo{author}{\bibfnamefont{S.~W.} \bibnamefont{Tozer}}, \bibnamefont{and}
  \bibinfo{author}{\bibfnamefont{C.}~\bibnamefont{Petrovic}},
  \bibinfo{journal}{Phys. Rev. B} \textbf{\bibinfo{volume}{84}},
  \bibinfo{pages}{220401} (\bibinfo{year}{2011}{\natexlab{a}}).

\bibitem[{\citenamefont{Wang et~al.}(2011{\natexlab{b}})\citenamefont{Wang,
  Zhao, Yin, Kotliar, Kim, Aronson, and Morosan}}]{PhysRevB.84.064428}
\bibinfo{author}{\bibfnamefont{J.~K.} \bibnamefont{Wang}},
  \bibinfo{author}{\bibfnamefont{L.~L.} \bibnamefont{Zhao}},
  \bibinfo{author}{\bibfnamefont{Q.}~\bibnamefont{Yin}},
  \bibinfo{author}{\bibfnamefont{G.}~\bibnamefont{Kotliar}},
  \bibinfo{author}{\bibfnamefont{M.~S.} \bibnamefont{Kim}},
  \bibinfo{author}{\bibfnamefont{M.~C.} \bibnamefont{Aronson}},
  \bibnamefont{and} \bibinfo{author}{\bibfnamefont{E.}~\bibnamefont{Morosan}},
  \bibinfo{journal}{Phys. Rev. B} \textbf{\bibinfo{volume}{84}},
  \bibinfo{pages}{064428} (\bibinfo{year}{2011}{\natexlab{b}}).

\bibitem[{\citenamefont{Guo et~al.}(2014)\citenamefont{Guo, Princep, Zhang,
  Manuel, Khalyavin, Mazin, Shi, and Boothroyd}}]{PhysRevB.90.075120}
\bibinfo{author}{\bibfnamefont{Y.~F.} \bibnamefont{Guo}},
  \bibinfo{author}{\bibfnamefont{A.~J.} \bibnamefont{Princep}},
  \bibinfo{author}{\bibfnamefont{X.}~\bibnamefont{Zhang}},
  \bibinfo{author}{\bibfnamefont{P.}~\bibnamefont{Manuel}},
  \bibinfo{author}{\bibfnamefont{D.}~\bibnamefont{Khalyavin}},
  \bibinfo{author}{\bibfnamefont{I.~I.} \bibnamefont{Mazin}},
  \bibinfo{author}{\bibfnamefont{Y.~G.} \bibnamefont{Shi}}, \bibnamefont{and}
  \bibinfo{author}{\bibfnamefont{A.~T.} \bibnamefont{Boothroyd}},
  \bibinfo{journal}{Phys. Rev. B} \textbf{\bibinfo{volume}{90}},
  \bibinfo{pages}{075120} (\bibinfo{year}{2014}).

\bibitem[{\citenamefont{Masuda et~al.}(2016)\citenamefont{Masuda, Sakai,
  Tokunaga, Yamasaki, Miyake, Shiogai, Nakamura, Awaji, Tsukazaki, Nakao
  et~al.}}]{masuda2016quantum}
\bibinfo{author}{\bibfnamefont{H.}~\bibnamefont{Masuda}},
  \bibinfo{author}{\bibfnamefont{H.}~\bibnamefont{Sakai}},
  \bibinfo{author}{\bibfnamefont{M.}~\bibnamefont{Tokunaga}},
  \bibinfo{author}{\bibfnamefont{Y.}~\bibnamefont{Yamasaki}},
  \bibinfo{author}{\bibfnamefont{A.}~\bibnamefont{Miyake}},
  \bibinfo{author}{\bibfnamefont{J.}~\bibnamefont{Shiogai}},
  \bibinfo{author}{\bibfnamefont{S.}~\bibnamefont{Nakamura}},
  \bibinfo{author}{\bibfnamefont{S.}~\bibnamefont{Awaji}},
  \bibinfo{author}{\bibfnamefont{A.}~\bibnamefont{Tsukazaki}},
  \bibinfo{author}{\bibfnamefont{H.}~\bibnamefont{Nakao}},
  \bibnamefont{et~al.}, \bibinfo{journal}{Sci. Adv.}
  \textbf{\bibinfo{volume}{2}}, \bibinfo{pages}{e1501117}
  (\bibinfo{year}{2016}).

\bibitem[{\citenamefont{May et~al.}(2014)\citenamefont{May, McGuire, and
  Sales}}]{PhysRevB.90.075109}
\bibinfo{author}{\bibfnamefont{A.~F.} \bibnamefont{May}},
  \bibinfo{author}{\bibfnamefont{M.~A.} \bibnamefont{McGuire}},
  \bibnamefont{and} \bibinfo{author}{\bibfnamefont{B.~C.} \bibnamefont{Sales}},
  \bibinfo{journal}{Phys. Rev. B} \textbf{\bibinfo{volume}{90}},
  \bibinfo{pages}{075109} (\bibinfo{year}{2014}).

\bibitem[{\citenamefont{Wang et~al.}(2013)\citenamefont{Wang, Graf, and
  Petrovic}}]{PhysRevB.87.235101}
\bibinfo{author}{\bibfnamefont{K.}~\bibnamefont{Wang}},
  \bibinfo{author}{\bibfnamefont{D.}~\bibnamefont{Graf}}, \bibnamefont{and}
  \bibinfo{author}{\bibfnamefont{C.}~\bibnamefont{Petrovic}},
  \bibinfo{journal}{Phys. Rev. B} \textbf{\bibinfo{volume}{87}},
  \bibinfo{pages}{235101} (\bibinfo{year}{2013}).

\bibitem[{\citenamefont{Borisenko et~al.}(2015)\citenamefont{Borisenko,
  Evtushinsky, Gibson, Yaresko, Kim, Ali, Buechner, Hoesch, and
  Cava}}]{borisenko2015time}
\bibinfo{author}{\bibfnamefont{S.}~\bibnamefont{Borisenko}},
  \bibinfo{author}{\bibfnamefont{D.}~\bibnamefont{Evtushinsky}},
  \bibinfo{author}{\bibfnamefont{Q.}~\bibnamefont{Gibson}},
  \bibinfo{author}{\bibfnamefont{A.}~\bibnamefont{Yaresko}},
  \bibinfo{author}{\bibfnamefont{T.}~\bibnamefont{Kim}},
  \bibinfo{author}{\bibfnamefont{M.}~\bibnamefont{Ali}},
  \bibinfo{author}{\bibfnamefont{B.}~\bibnamefont{Buechner}},
  \bibinfo{author}{\bibfnamefont{M.}~\bibnamefont{Hoesch}}, \bibnamefont{and}
  \bibinfo{author}{\bibfnamefont{R.~J.} \bibnamefont{Cava}},
  \bibinfo{journal}{arXiv preprint arXiv:1507.04847}  (\bibinfo{year}{2015}).

\bibitem[{\citenamefont{Liu et~al.}(2015)\citenamefont{Liu, Hu, Graf,
  Radmanesh, Adams, Zhu, Chen, Liu, Wei, Chiorescu et~al.}}]{liu2015discovery}
\bibinfo{author}{\bibfnamefont{J.}~\bibnamefont{Liu}},
  \bibinfo{author}{\bibfnamefont{J.}~\bibnamefont{Hu}},
  \bibinfo{author}{\bibfnamefont{D.}~\bibnamefont{Graf}},
  \bibinfo{author}{\bibfnamefont{S.}~\bibnamefont{Radmanesh}},
  \bibinfo{author}{\bibfnamefont{D.}~\bibnamefont{Adams}},
  \bibinfo{author}{\bibfnamefont{Y.}~\bibnamefont{Zhu}},
  \bibinfo{author}{\bibfnamefont{G.}~\bibnamefont{Chen}},
  \bibinfo{author}{\bibfnamefont{X.}~\bibnamefont{Liu}},
  \bibinfo{author}{\bibfnamefont{J.}~\bibnamefont{Wei}},
  \bibinfo{author}{\bibfnamefont{I.}~\bibnamefont{Chiorescu}},
  \bibnamefont{et~al.}, \bibinfo{journal}{arXiv preprint arXiv:1507.07978}
  (\bibinfo{year}{2015}).

\bibitem[{\citenamefont{Cordier and
  Sch{\"a}fer}(1977)}]{cordier1977darstellung}
\bibinfo{author}{\bibfnamefont{G.}~\bibnamefont{Cordier}} \bibnamefont{and}
  \bibinfo{author}{\bibfnamefont{H.}~\bibnamefont{Sch{\"a}fer}},
  \bibinfo{journal}{Z.Naturforsch.(B)} \textbf{\bibinfo{volume}{32}},
  \bibinfo{pages}{383} (\bibinfo{year}{1977}).

\bibitem[{\citenamefont{Zhang et~al.}(2009)\citenamefont{Zhang, Su, Lu, Weng,
  Lee, and Xiang}}]{zhang2009universal}
\bibinfo{author}{\bibfnamefont{G.-M.} \bibnamefont{Zhang}},
  \bibinfo{author}{\bibfnamefont{Y.-H.} \bibnamefont{Su}},
  \bibinfo{author}{\bibfnamefont{Z.-Y.} \bibnamefont{Lu}},
  \bibinfo{author}{\bibfnamefont{Z.-Y.} \bibnamefont{Weng}},
  \bibinfo{author}{\bibfnamefont{D.-H.} \bibnamefont{Lee}}, \bibnamefont{and}
  \bibinfo{author}{\bibfnamefont{T.}~\bibnamefont{Xiang}},
  \bibinfo{journal}{EPL} \textbf{\bibinfo{volume}{86}}, \bibinfo{pages}{37006}
  (\bibinfo{year}{2009}).

\bibitem[{\citenamefont{Abrikosov and
  Beknazarov}(1988)}]{abrikosov1988fundamentals}
\bibinfo{author}{\bibfnamefont{A.~A.} \bibnamefont{Abrikosov}}
  \bibnamefont{and}
  \bibinfo{author}{\bibfnamefont{A.}~\bibnamefont{Beknazarov}},
  \emph{\bibinfo{title}{Fundamentals of the Theory of Metals}}
  (\bibinfo{publisher}{North-Holland Amsterdam}, \bibinfo{year}{1988}).

\bibitem[{\citenamefont{Abrikosov}(1998)}]{PhysRevB.58.2788}
\bibinfo{author}{\bibfnamefont{A.~A.} \bibnamefont{Abrikosov}},
  \bibinfo{journal}{Phys. Rev. B} \textbf{\bibinfo{volume}{58}},
  \bibinfo{pages}{2788} (\bibinfo{year}{1998}).

\bibitem[{\citenamefont{Abrikosov}(2000)}]{abrikosov2000quantum}
\bibinfo{author}{\bibfnamefont{A.}~\bibnamefont{Abrikosov}},
  \bibinfo{journal}{EPL} \textbf{\bibinfo{volume}{49}}, \bibinfo{pages}{789}
  (\bibinfo{year}{2000}).

\bibitem[{\citenamefont{Hu and Rosenbaum}(2008)}]{hu2008classical}
\bibinfo{author}{\bibfnamefont{J.}~\bibnamefont{Hu}} \bibnamefont{and}
  \bibinfo{author}{\bibfnamefont{T.}~\bibnamefont{Rosenbaum}},
  \bibinfo{journal}{Nat. Mater.} \textbf{\bibinfo{volume}{7}},
  \bibinfo{pages}{697} (\bibinfo{year}{2008}).

\bibitem[{\citenamefont{Lee et~al.}(2002)\citenamefont{Lee, Rosenbaum,
  Saboungi, and Schnyders}}]{PhysRevLett.88.066602}
\bibinfo{author}{\bibfnamefont{M.}~\bibnamefont{Lee}},
  \bibinfo{author}{\bibfnamefont{T.~F.} \bibnamefont{Rosenbaum}},
  \bibinfo{author}{\bibfnamefont{M.-L.} \bibnamefont{Saboungi}},
  \bibnamefont{and} \bibinfo{author}{\bibfnamefont{H.~S.}
  \bibnamefont{Schnyders}}, \bibinfo{journal}{Phys. Rev. Lett.}
  \textbf{\bibinfo{volume}{88}}, \bibinfo{pages}{066602}
  (\bibinfo{year}{2002}).

\bibitem[{\citenamefont{Huynh et~al.}(2011)\citenamefont{Huynh, Tanabe, and
  Tanigaki}}]{PhysRevLett.106.217004}
\bibinfo{author}{\bibfnamefont{K.~K.} \bibnamefont{Huynh}},
  \bibinfo{author}{\bibfnamefont{Y.}~\bibnamefont{Tanabe}}, \bibnamefont{and}
  \bibinfo{author}{\bibfnamefont{K.}~\bibnamefont{Tanigaki}},
  \bibinfo{journal}{Phys. Rev. Lett.} \textbf{\bibinfo{volume}{106}},
  \bibinfo{pages}{217004} (\bibinfo{year}{2011}).

\bibitem[{\citenamefont{Li et~al.}(2016)\citenamefont{Li, Wang, Graf, Wang,
  Wang, and Petrovic}}]{Petrovic2016}
\bibinfo{author}{\bibfnamefont{L.}~\bibnamefont{Li}},
  \bibinfo{author}{\bibfnamefont{K.}~\bibnamefont{Wang}},
  \bibinfo{author}{\bibfnamefont{D.}~\bibnamefont{Graf}},
  \bibinfo{author}{\bibfnamefont{L.}~\bibnamefont{Wang}},
  \bibinfo{author}{\bibfnamefont{A.}~\bibnamefont{Wang}}, \bibnamefont{and}
  \bibinfo{author}{\bibfnamefont{C.}~\bibnamefont{Petrovic}},
  \bibinfo{journal}{Phys. Rev. B} \textbf{\bibinfo{volume}{93}},
  \bibinfo{pages}{115141} (\bibinfo{year}{2016}).

\end{thebibliography}
\end{document}